\documentclass[twocolumn,showpacs,preprintnumbers,amsmath,amssymb,superscriptaddress]{revtex4}

\usepackage{graphicx}
\usepackage{dcolumn}
\usepackage{bm}

\begin{document}

\title{Laboratory measurements of electrostatic solitary structures\\ generated by electron beam injection\footnote{{\it Laboratory measurements of electrostatic solitary structures generated by beam injection}, Bertrand Lefebvre, Li-Jen Chen, Walter Gekelman, Paul Kintner, Jolene Pickett, Patrick Pribyl, Stephen Vincena, Franklin Chiang, Jack Judy, Physical Review Letters, {\bf 105}, 115001, doi:10.1103/PhysRevLett.105.115001, 2010.
 Copyright (2010) by the American Physical Society.}}

\author{Bertrand Lefebvre}
\affiliation{Space Science Center, University of New Hampshire, Durham, New Hamphire 03824, USA}
\email{bertrand.lefebvre@unh.edu}
\author{Li-Jen Chen}
\affiliation{Space Science Center, University of New Hampshire, Durham, New Hamphire 03824, USA}
\author{Walter Gekelman}
\affiliation{Basic Plasma Science Facility, University of California, Los Angeles, California 90095, USA}
\author{Paul Kintner}
\affiliation{School of Electrical and Computer Engineering, Cornell University, Ithaca, New York 14853, USA}
\author{Jolene Pickett}
\affiliation{Department of Physics and Astronomy, University of Iowa, Iowa City, Iowa 52242, USA}
\author{Patrick Pribyl}
\affiliation{Basic Plasma Science Facility, University of California, Los Angeles, California 90095, USA}
\author{Stephen Vincena}
\affiliation{Basic Plasma Science Facility, University of California, Los Angeles, California 90095, USA}
\author{Franklin Chiang}
\affiliation{Electrical Engineering Department, University of California, Los Angeles, California 90095, USA}
\author{Jack Judy}
\affiliation{Electrical Engineering Department, University of California, Los Angeles, California 90095, USA}

\date{\today}

\begin{abstract}
Electrostatic solitary structures are generated by injection of a suprathermal electron beam parallel to the magnetic field in a laboratory plasma. Electric microprobes with tips smaller than the Debye length ($\lambda_{De}$) enabled the measurement of positive potential pulses with half-widths 4 to 25$\lambda_{De}$ and velocities 1 to 3 times the background electron thermal speed. Nonlinear wave packets of similar velocities and scales are also observed, indicating that the two descend from the same mode which is consistent with the electrostatic whistler mode and result from an instability likely to be driven by field-aligned currents.
\end{abstract}


\maketitle


Electrostatic solitary structures identified as Bernstein-Greene-Kruskal electron holes \cite{Tur84PS} have been abundantly observed in various active regions of
space subject to large-scale current systems or energy dissipation (\cite{ECM98PRL,FKP05JGR} and references therein). This includes regions where magnetic reconnection occurs \cite{DSC03Sci,MDK03GRL}. Electron holes were also recently detected in a laboratory magnetic reconnection experiment \cite{FPE08PRL}. Electron holes can be generated by energetic streaming electrons and are thought to play an important role in scattering these electrons. However, although these small-scale (one to tens of Debye lengths, $\lambda_{De}$) solitary structures seem ubiquitous in key regions of space their exact origin  often remains unclear. In the laboratory, experiments dedicated to electron holes were carried out in a strongly magnetized Q-machine  \cite{SMP79PRL}. These holes were generated by a voltage pulse and had sizes comparable to the plasma column radius, making comparison with holes observed in space difficult.

\begin{figure}
  \noindent\centering
  \includegraphics[width=0.98\columnwidth]{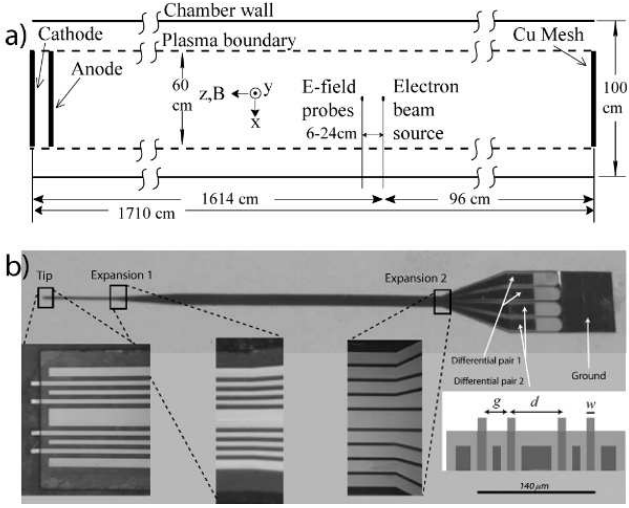}
  \caption{a) Schematic of probe and small electron beam in the LAPD device. b) Highly magnified photograph of the electric field MEMS microprobe. The conductor is made from gold and the insulating material is polyimide.  The dimensions are w=10 $\mu$m, g = 30 $\mu$m and d= 70 $\mu$m. The wires are coaxial and tapered in their expansion regions to maintain a 50 $\Omega$ impedance \cite{Chi09UCLA}.}
  \label{fig:sketch}
\end{figure}

This Letter reports measurements of electrostatic solitary waves generated by an electron beam injected into a magnetized low-$\beta$ plasma column much larger than the structure scales (Fig. \ref{fig:sketch}). The experiment was conducted at the upgraded Large Plasma Device (LAPD) \cite{GPL91RSI} at the University of California, Los Angeles. The helium plasma column has a 60 cm diameter, is 17.1 m long and pulsed at 1 Hz with pulses lasting several milliseconds (Fig. \ref{fig:sketch}a). An electron beam 0.4 to 1 cm in diameter is injected from a 3 mm diameter LaB6 crystal source for about 140 $\mu$s along the axis of the column in the afterglow phase, between 50 and 150 ms after the end of the discharge pulse. The beam density 5 cm from the source is approximately 25\% of the background electron density. The magnetic field strength, plasma density and beam voltage can be changed from experiment to experiment. The range of the main plasma parameters is summarized in table \ref{tab:params}. Floating potentials and electric fields are measured at 20 GHz (38$f_{pe}$ for the average plasma density) by a micro-probe \cite{Chi09UCLA} with 10 $\mu$m wide tips (at most $\lambda_{De}$/4) and with separations ranging from 40 to 
130 $\mu$m ($\sim\lambda_{De}$) [Fig. \ref{fig:sketch}b]. Microprobe tips smaller than $\lambda_{De}$ are critical to study Debye-scale structures without destroying them. The probe is located on the column central axis, 6--24 cm or 150--5600$\lambda_{De}$ away from the beam source and parallel to the magnetic field and the column axis.

\begin{table}
  \caption{Main plasma parameter range in the afterglow phase during which the measurements are taken. Magnetic field strength and beam energy are fixed parameters for each experiment, while densities and temperatures change with time in a highly reproducible way.}
  \label{tab:params}
  \begin{tabular}{ll}
  \tableline
  Magnetic field strength, $B$              &  100--750 G \\
  Background electron density, $n_e$        &  1.1--5.7$\times 10^9$ cm$^{-3}$ \\
  Background electron temperature, $T_e$    &  0.18--0.20 eV \\
  Electron beam voltage                     &  60--120  V \\
  Debye length, $\lambda_{De}$              &  44--95 $\mu$m \\
  Electron mean-free path, $\lambda_{ee}$   &  0.11--0.42 m  \\
  Thermal electron gyroradius, $r_{ge}$     &  13--106 $\mu$m \\
  Plasma frequency, $f_{pe}$                &  297--681 MHz\\
  Electron gyrofrequency, $f_{ce}$          &  0.53--7$f_{pe}$ \\  
  \tableline\\
  \end{tabular}
\end{table}

\begin{figure}
  \noindent\centering\includegraphics[width=0.8\columnwidth]{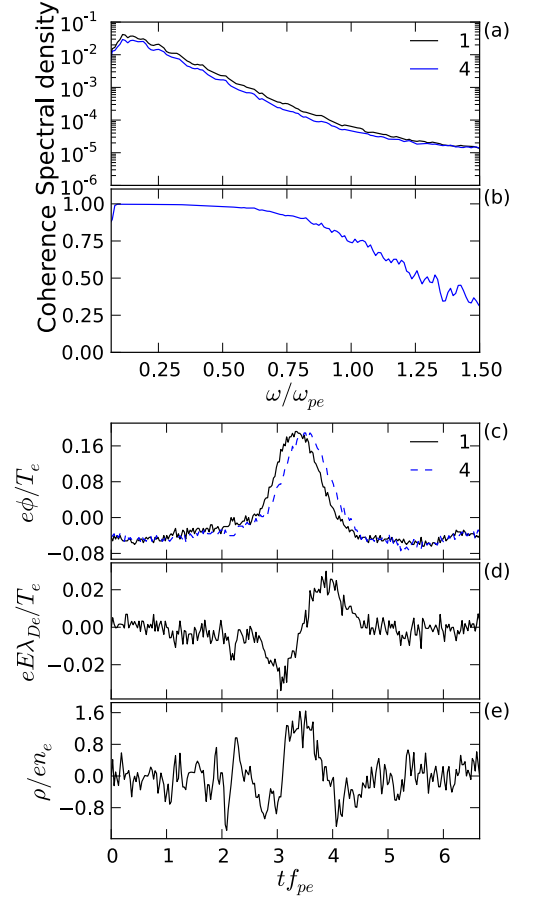}
  \caption{Power spectrum (a) and coherence (b) for potential time-series on two field-aligned probes separated by 2.3$\lambda_{De}$. A solitary potential pulse (c), its bipolar electric field (d), and estimated charge density perturbation (e). In this and all other figures the parameters used for normalization correspond to the unperturbed state before beam injection.}
  \label{fig:spectrum_hole}
\end{figure}

The fluctuations excited by the beam typically have a broadband spectrum ranging from the bottom of the amplifier's frequency range (20 MHz) to the electron plasma frequency $f_{pe}$ and above. For the example shown in Fig. \ref{fig:spectrum_hole}a the power-spectrum appears exponential for $0.2<f/f_{pe}<0.8$. Fluctuations in this range display a high degree of phase coherence between signals measured on probes separated by 2.3$\lambda_{De}$ (Fig. \ref{fig:spectrum_hole}b), and consist of a mixture of quasiperiodic waves, wave packets and solitary pulses. Solitary waves alone are also known in some cases to result in exponential spectra \cite{PSM08POP}.

An example of a solitary positive potential pulse is shown in Figs. \ref{fig:spectrum_hole}cde. The figure displays the potentials recorded on two probe tips separated by $\Delta=1.77\lambda_{De}$ (100 $\mu$m) along the magnetic field direction, the electric field inferred as $E=-\delta\phi/\Delta$ and a lower bound (in magnitude) to the actual density perturbation calculated as $\rho=\varepsilon_0 dE/dx$ assuming a slablike structure propagating at a constant velocity. Quantities are made dimensionless using the plasma parameters in the unperturbed state before beam injection. In this example $n_e=3.5\times10^{9}$ cm$^{-3}$, $T_e=0.2$ eV and $B=750$ G ($f_{pe}/f_{ce} = 1/4$). The pulse conserves its shape while traveling between the probe tips. The perturbation is small in the sense that $e\phi<T_e$ and $\Delta n_e\ll n_e$. The time delay between probes is 0.2$f_{pe}^{-1}$ (0.3 ns), implying a parallel velocity in the same direction as the beam electrons $v_{\parallel}=1.98v_{Te}=371$ km/s with $v_{Te}=(T_e/m_e)^{1/2}$, and a parallel half-width $L_{\parallel}=10.3\lambda_{De}$ (575 $\mu$m). The pulse has a symmetric positive potential with dipolar electric field and a charge density which is positive at the center and negative near the edges. These properties are consistent with a Bernstein-Greene-Kruskal electron hole \cite{Tur84PS}.

\begin{figure}
  \noindent\centering
  \includegraphics[width=0.8\columnwidth]{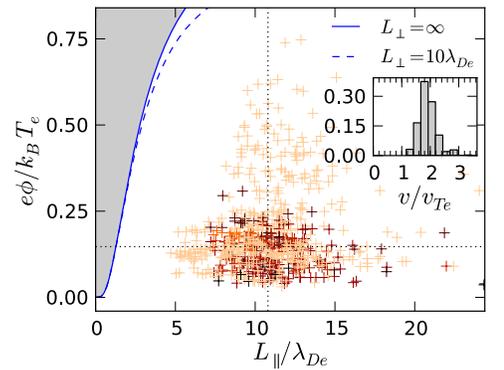}
\caption{Scatter plots of solitary pulse half-width versus amplitude. The shaded area corresponds to a theoretically inaccessible region \cite{CPK05JGR}. Points are color coded according to the magnetic field ranging from 100 G (black/dark red) to 750 G (light orange). The dotted lines indicate the median values. The inset shows a probability histogram of the structure velocities.}
  \label{fig:stats}
\end{figure}

An automated search for isolated peaks in the electric potential time series for plasma parameters in Table \ref{tab:params} yielded a total of 363 solitary structures over 35 experiments with various beam voltages, plasma densities and magnetic fields. The propagation time from one probe tip to another is determined by cross-correlation analysis, and velocities cross-checked between different pairs of probe tips. All structures propagate in the same direction as the beam and display a main positive potential peak. 40\% have symmetric dipolar electric fields, while 30\% are more tripolar and the remainder mostly dipolar with a notable asymmetry. Scatter plots of their main parameters are shown in Fig. \ref{fig:stats}. The half-widths range from 4.5 to 24.4$\lambda_{De}$ with a median of 10.2$\lambda_{De}$. Amplitudes are all smaller than $T_e/e$ with a median of 0.13$T_e/e$. 80\% of the structures have velocities between 1.3 and 2.3$v_{Te}$ (Fig. \ref{fig:stats}, inset).

The typical scales are comparable to those usually reported in space observations, albeit a bit larger. References \cite{ECM98PRL,FKP05JGR} report average half-width slightly larger than $\lambda_{De}$ and velocity slightly smaller than $v_{Te}$. This difference might be partly explained by the normalization, which in our cases uses background parameters measured prior to the beam injection. In particular the plasma in which the solitary waves propagate is likely warmer due to the interaction with the beam, making the structure velocities closer to the actual thermal speed. Indeed it was predicted that electron holes faster than $2v_{Te}$ are unstable \cite{Tur84PS} and are therefore unlikely to be observed. Normalization questions apart, electron hole sizes significantly larger than $\lambda_{De}$ are theoretically allowed. electron holes of parallel size $60\lambda_{De}$ have been reported in a laboratory experiment \cite{FPE08PRL}.

The widths and amplitudes do not display any one-to-one relation (Fig. \ref{fig:stats}), unlike Korteweg de Vries-type solitons. However the positivity of the trapped electron distribution imposes an upper limit to the hole amplitude for a given scale \cite{CPK05JGR}. All the amplitudes lie well within the allowed region for stationary 1d (infinite perpendicular size $L_{\perp}$) or magnetized 3d (with finite $L_{\perp}$) electron holes predicted by \cite{CPK05JGR}. The electron trapping frequency $f_{tr}=(ek_{\parallel}|E_{\parallel}|/m_e)^{1/2}$ is found to be  typically a few percent of $f_{ce}$, showing that electrons remain magnetized within these structures. However parallel sizes or the amplitudes have no clear relationship to the magnetic field intensity (color-coded in Fig.  \ref{fig:stats}).

\begin{figure}
  \noindent\centering
  \includegraphics[width=0.8\columnwidth]{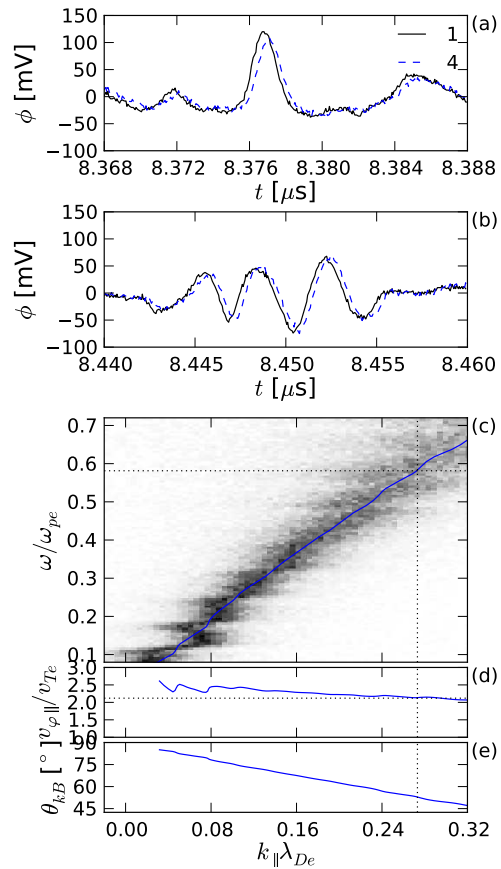}
  \caption{(a) potentials on two probes showing a solitary structure. (b) a wave packet shortly following the previous solitary structure. (c) experimental dispersion relation $\omega(k_{\parallel})$ (the solid blue curve is the average). (d) phase velocity derived from the experimental dispersion relation. (e) $\theta_{kB}$ derived from Eq. \ref{eq:disp}. The horizontal dotted curve in (c) is the inverse median time-scale of the electron holes in this particular experiment, and the one in (d) is their median velocity. The vertical dotted line in (c)-(e) is the wave-number corresponding to their median parallel width. In this experiment $f_{ce}/f_{pe}=3$.}
  \label{fig:disp}
\end{figure}

Besides electron holes, irregular fluctuations and wave packets are observed. Fig. \ref{fig:disp}ab shows an example of a solitary structure shortly followed by a wave packet. The approximately constant phase-shift shows that the solitary wave and wave packet have very similar velocities and scales and therefore might be related to the same plasma mode. Measurements on different probes allow estimation of the local wave-number and spectral density based on the instantaneous cross-spectra \cite{bkp92jap}. The density and the average experimental dispersion relation $\omega(k_{\parallel})$ are shown in Fig. \ref{fig:disp}c. The fluctuations follow a weakly dispersive curve, although the spectral density becomes smaller and more spread out as the frequency increases toward $f_{pe}$. The average experimental dispersion curve extends to the temporal and spatial scales where the electron holes are observed and implies parallel phase (and group) velocities about $2v_{Te}$, very similar to the measured velocities of the holes (their median velocity for these experimental parameters is 2.1$v_{Te}$, the horizontal dotted line in Fig 4d). In this frequency and wave-number range the dispersion is consistent with the whistler mode at the resonance angle, which is essentially electrostatic and has a lower cutoff at the lower-hybrid frequency $\omega_{LH}$. In the frequency and wave-number range we observe the waves ion and beam contributions can be neglected and their dispersion relation reduces to 
\begin{equation}
\frac{k_{\perp}^2}{k_{\parallel}^2} = -\frac{P}{S} \approx - \frac{1 - \frac{\omega_{pe}^2}{\omega^2}}{1 - \frac{\omega_{pe}^2}{\omega^2 - \Omega_{ce}^2}} \label{eq:disp}
\end{equation}
where P and S are standard elements of the cold plasma dielectric tensor \cite{swanson}. Electrostatic whistler mode waves following the average experimental curve $\omega(k_{\parallel})$ have an angle predicted by Eq. \ref{eq:disp} shown in Fig. \ref{fig:disp}e. The waves are predicted to become more perpendicular as $k_{\parallel}$ decreases and have $k_{\perp}\lambda_{De}$ between 0.33 and 0.39. Electron holes and electrostatic whistler mode waves known as VLF saucers have been found to be closely associated in the auroral regions \cite{ECM01GRL}. They have also been observed together in the polar cusp \cite{Fra00Cornell,PFS01JGR}. The close association between the electron holes and electrostatic whistler mode waves suggests that both originate from the same source, or that one produces the other \cite{ONG99PRL}.

Knowledge of the electron distribution function is key in sorting out the type of instability that is responsible for the generation of the observed waves and perhaps the solitary structures.  The beam distribution was measured using a swept Langmuir probe capable of 100 V sweeps. The derivative of the I-V characteristic curve is proportional to the distribution function.  32 cm away along the background magnetic field from the 66 V beam source no beam is visible, and only a 35 eV tail is left.  However at the probe location where most of the solitary structures are measured (about 5-6 cm from the beam source), a warm beam with a mean energy of 62 eV, thermal energy $T_b\approx 5$ eV and a density of approximately 25\% of the total density is observed. In addition a high energy tail on the background distribution extending up to the beam energy is observed with a 30\% relative density. This tail must result from an instability occurring closer to the beam source.

The low velocities of the electron holes ($0.1$--0.2$v_{\mathrm{beam}}$) and their lack of dependence on the beam velocity excludes generation by a two-stream instability, which here predicts a velocity $\approx 2v_{\mathrm{beam}}/3$. The two-stream instability generation mechanism was discarded for analogous reasons by \cite{DSC03Sci} and \cite{CDW05JGR}. Similarly the electron holes or electrostatic whistler mode waves in this frequency range cannot be directly generated by a resonant ($\omega - k_{\parallel}v_{\mathrm{beam}}= 0$) interaction with beam electrons or a fan instability based on anomalous Doppler-shifted resonance ($\omega - k_{\parallel}v_{\parallel}=-\omega_{ce}$) with electrons from the tail. However the observed electron distribution carries a substantial parallel current. Parallel currents in low-$\beta$ plasmas  can drive unstable oblique electrostatic whistlers up  to the frequency range of our measurements \cite{MC06POP} (for case b in this article). In our experiment a 62 eV 25\% beam results in a mean electron velocity $\approx 4.4 v_{Te}$, in which case this instability predicts a parallel phase velocity of $ 2.2 v_{Te}$, much closer to the measurements. Note however that the observed electron distribution has a background electron population not considered in \cite{MC06POP}. It also remains unclear if this instability can directly generate electron holes or they form later as the waves travel away from the beam center and the region of instability.

In summary, for the first time a laboratory experiment has shown the generation of localized electrostatic structures by the injection of a suprathermal electron beam. The scales and amplitudes of these structures are comparable to those derived from observations in various places in the magnetosphere, and their properties are consistent with electron holes. The electron holes are found on the high frequency ($f\leq f_{pe}$) end of the electrostatic whistler mode which suggests that they have the same origin. We conjecture they result from a lower-hybrid instability driven by parallel currents.

The work at UNH was supported by DOE under grant DE-FG02-07ER54941, at Cornell by DOE DE-FG02-07ER54942, and at UI by DOE grant DE-FG02-07ER54943 as well as by NASA Goddard Space Flight Center through NNX07AI24G. The Work at UCLA was done at the Basic Plasma Science Facility, BaPSF, which is supported by the NSF PHY-0531621 and DOE DE-FC02-07ER54918.

\end{document}